\documentclass[twocolumn,prd,aps,floats,nofootinbib,tightenlines,showpacs,floatfix]{revtex4-1}
\usepackage{graphicx}
\usepackage{amsmath}
\usepackage{amsfonts}
\usepackage{amssymb,amsthm}
\usepackage{mathrsfs}
\usepackage{epsfig}
\usepackage{color}
\begin{document}
\title{Cavitation and thermal photon production in relativistic heavy ion collisions}
\author{Jitesh R. Bhatt}
\email{jeet@prl.res.in}
\author{Hiranmaya Mishra}
\email{hm@prl.res.in}
\author{V. Sreekanth}
\email{skv@prl.res.in}
\affiliation{Physical Research Laboratory, Ahmedabad 380009, India}
\begin{abstract}
 We investigate the thermal photon production-rates using one dimensional boost-invariant second order relativistic 
hydrodynamics to find  proper time evolution of the energy density and the temperature. The effect of bulk-viscosity and
non-ideal equation of state are taken into account in a manner consistent with recent lattice QCD estimates. 
It is shown that the \textit{non-ideal} gas equation of state i.e $\varepsilon-3\,P\,\neq 0$ behaviour of the expanding plasma,
which is important near the phase-transition point, can significantly slow down the hydrodynamic expansion and 
thereby increase the photon production-rates. Inclusion of the bulk viscosity may also have similar effect on the hydrodynamic evolution. 
However the effect of bulk viscosity is shown to be significantly lower than the \textit{non-ideal} gas equation of state.  
We also analyze the interesting phenomenon of bulk viscosity induced cavitation making the hydrodynamical description
invalid. It is shown that ignoring the cavitiation phenomenon can lead to a very significant over estimation of the photon flux.
It is argued that this feature could be relevant in studying signature of cavitation in relativistic heavy ion collisions.

\end{abstract}
\maketitle
\section{INTRODUCTION}

Thermal photons emitted from the hot fireball created in relativistic heavy-ion
collisions is a promising tool for providing a signature of quark-gluon plasma
\cite{kapu91,baier92,Ruus92,thoma95,traxler95,arnoldJHEP} 
(see \cite{Alam01R,Peitz-Thoma:02R,Gale:03R} for recent reviews). 
Since they participate only in electromagnetic 
interactions, they 
have a larger mean free path compared to the transverse size of the hot and dense matter created 
in nuclear collisions 
\cite{kapusta}. 
Therefore  these photons were proposed to verify the existence of the QGP phase 
\cite{Feinberg76,Shyryak78}.
Spectra of thermal photons depend upon the
fireball temperature and they can be calculated from the scattering cross-section of
the processes like $q\bar q\rightarrow g\gamma$, \textit{bremsstrahlung} etc. Time evolution
of the temperature can be calculated using hydrodynamics with appropriate
initial conditions. 
Thus the spectra depend upon the equation of state (EoS) of the medium and they 
may be useful in finding a signature of the quark-gluon plasma\cite{dk99:EPJC,thoma07,dkdir08,Liu09}. 
Recently the thermal photons are proposed as a tool to measure the shear viscosity
of the strongly interacting matter produced in the collisions\cite{skv-photon,dusling09}.

 Understanding shear viscosity of QGP is one of the most
intriguing aspects of the experiments at Relativistic Heavy Ion Collider (RHIC).
Analysis of the experimental data collected from RHIC show
that the strongly coupled matter produced in the collisions is not too much above
the phase transition temperature $T_c$ and it may have extremely small value of shear viscosity
$\eta$. In fact the ratio of the shear viscosity
$\eta$ to the entropy density $s$ i.e. $\eta/s$ is around $1/4\pi$ which is the smallest for
any known liquid in the nature\cite{Schf-Teaney-09R}. 
In fact the arguments based on AdS-CFT suggest that 
the values of $\eta/s$ can not become lower than $1/4\pi$. This is now 
known as  Kovtun-Son-Starinets or 'KSS- bound' \cite{kss05}. Thus the quark-gluon plasma
produced in RHIC experiments is believed to be in a form of the most perfect liquid\cite{Hirano:2005wx}.
No wonder ideal hydrodynamic appears to be best description of such matter as suggested by
comparison between the experimental data\cite{RHIC} and the calculations done using
second-order relativistic hydrodynamics
\cite{BR07,rom:PRL,DT08,hs08,LR08,Molnar08,Song:2008hj,LR09}.

However there remains uncertainties in understanding the application and validity
of the hydrodynamical procedure in the relativistic heavy-ion collision experiments. 
It is only very recently realized that the effect of bulk viscosity can bring complications
in the hydrodynamical description of the heavy-ion collisions. 
Generally it was believed that the bulk viscosity does not play a significant
role in the hydrodynamics of relativistic heavy-ion collisions. 
It was argued that that since $\zeta$ scaled like  $\varepsilon -3P$ 
at very high energy the bulk viscosity may not play a significant role because
the matter might be following the ideal gas type equation of state\cite{wein}. But during its course
of expansion the fireball temperature can  approach values close to $T_c$. 
Recent lattice QCD results may not have ideal gas EoS 
and the ratio $\zeta/s$ show a strong peak around $T_c$ \cite{Bazavov:2009,Meyer:2008}. The
bulk viscosity contribution in this regime can be much larger than that of the the shear viscosity.
Recently the role of bulk viscosity in heating and expansion of the fireball was analyzed
using one dimensional hydrodynamics\cite{fms08}. 
Another complication that bulk viscosity brings in hydrodynamics of
heavy-ion collisions is phenomenon of cavitation\cite{kr:2010cv}. Cavitation arises
when the fluid pressure becomes smaller than the vapour pressure. Since the
bulk viscosity (and also shear viscosity) contributes to the pressure gradient
with a negative contribution, it may be possible for the effective fluid pressure
to become zero. Once the cavitation sets in the hydrodynamical description
breaks down. It was shown in Ref.\cite{kr:2010cv} that cavitation may happen in
RHIC experiments when the effect of bulk viscosity is included in manner consistent
with the lattice results. It was shown that the cavitation may significantly
reduce the time of hydrodynamical evolution. 

One of the main objectives of this paper is to study the photon spectra
with the effect of bulk viscosity and cavitation. Finite $\zeta$ effect can either 
significantly reduce the time for the hydrodynamical evolution (by onset of cavitation)
or  it can increase the time by which the system reaches $T_c$! Moreover the \textit{non-ideal}
gas EoS can also significantly influence the hydrodynamics (see below). 
In what follows, we use equations of relativistic second order hydrodynamics to
incorporate the effects of finite viscosity. We take the value of $\zeta/s$ same
as that in Ref.\cite{kr:2010cv} and keep $\eta/s=1/4\pi$. Further we use one dimensional
boost invariant hydrodynamics in the same spirit as in Refs.\cite{fms08,kr:2010cv}.
One of the limitations of this approach is that the effects of transverse flow cannot
be incorporated. As the boost-invariant hydrodynamics is known to lead to underestimation
of the effects of bulk viscosity\cite{fms08}, we believe that our study of the photon
spectra will provide a conservative estimate of the effect. 

\section{FORMALISM}
\subsection{Viscous Hydrodynamics}
We represent the energy momentum tensor of the dissipative QGP formed in high energy nuclear collisions as
\begin{equation}
T^{\mu\nu}=\varepsilon \, u^\mu\,u^\nu\, - P\, \Delta^{\mu\nu} + \Pi^{\mu\nu}
\label{Tmunu}
\end{equation}
where $\varepsilon$, $P$ and $u^\mu$ are the energy density, pressure and four velocity of the fluid element respectively. 
The operator $\Delta^{\mu\nu}~=~g^{\mu\nu} - u^\mu\,u^\nu$ acts as a projection perpendicular to four velocity. 
The viscous contributions to $T^{\mu\nu}$ are represented by 
\begin{equation}
 \Pi^{\mu\nu} = \pi^{\mu\nu}-\,\Delta^{\mu\nu}\, \Pi
\end{equation}
where $\pi^{\mu\nu}$, the traceless part of $\Pi^{\mu\nu}$; gives the contribution of shear viscosity and $\Pi$ gives the 
bulk contribution. The corresponding equations of motion are given by,
\begin{eqnarray}
D \varepsilon + (\varepsilon+P)\, \theta-\Pi^{\mu\nu}\nabla_{(\mu}\,u_{\nu)}=0\label{edot}\\
(\varepsilon+P)\,Du^{\alpha}-\nabla^{\alpha}P\,+\,\Delta_{\alpha\nu}\,\partial_\mu \Pi^{\mu\nu}=0
\end{eqnarray}
where $D\equiv u^\mu\partial_\mu$, $\theta\equiv\partial_{\mu}\,u^\mu$, $\nabla_{\alpha}=\Delta_{\mu\alpha}\partial^{\mu}$ 
and $A_{(\mu}\,B_{\nu)}=\frac{1}{2}
[A_\mu\,B_\nu+A_\nu\,B_\mu]$ gives the symmetrization. \\

We employ Bjorken's prescription\cite{bjor} to describe the one dimensional boost invariant expanding flow, 
were we use the convenient parametrization of the coordinates using the proper time $\tau = \sqrt{t^2-z^2}$ and space-time 
rapidity $y=\frac{1}{2}\,ln[\frac{t+z}{t-z}]$; $t=\tau$ cosh$\,y$ and $z=\tau$ sinh$\,y$. Then the four velocity is given by,
\begin{equation}
u^\mu=(\rm{cosh}\,y,0,0,\rm{sinh}\,y).
\end{equation}
We note that with this transformation of the coordinates, $D=\frac{\partial\,}{\partial\tau}$ and $\theta=1/\tau$. 

Form of the energy momentum tensor in the local rest frame 
of the fireball is then given by\cite{Teaney03,SHC06,AM07,AM-prl02} 
\begin{equation}
T^{\mu\nu} = \left(
\begin{array}{cccc}
\varepsilon & 0 & 0 & 0 \\
0 & P_{\perp} & 0 & 0 \\
0 & 0 & P_{\perp} & 0 \\
0 & 0 & 0 & P_{z} 
\end{array} \right) 
\label{Tmunu}
\end{equation}
where the effective pressure of the expanding fluid 
in the transverse and longitudinal directions are respectively given by 
\begin{eqnarray}
P_{\perp} &=& P + \Pi + \frac{1}{2}  \Phi
\nonumber \\
P_{z} &=& P + \Pi - \Phi
\label{pressures}
\end{eqnarray}
Here $\Phi$ and $\Pi$ are the non-equilibrium contributions to the  equilibrium pressure $P$
coming from shear and bulk viscosities. Respecting the symmetries in the 
transverse directions the traceless
shear tensor has the form $\pi^{ij} = \mathrm{diag}(\Phi/2, \Phi/2,-\Phi)$.

In the first order Navier-Stokes dissipative hydrodynamics 
\begin{equation}
  \label{NS}
  \Pi = -\zeta \partial_\mu u^\mu \quad \rm{and} \quad \pi^{\mu\nu} = 2\eta 
   \nabla^{\langle\mu} u^{\nu\rangle} \, ,
\end{equation}
with $\zeta,\eta>0$ and $\nabla_{\langle\mu} u_{\nu\rangle}=2\nabla_{(\mu}\,u_{\nu)}
-\frac{2}{3}\,\Delta_{\mu\nu}\nabla_\alpha u^\alpha$. So for first order theories with Bjorken flow we have
\begin{equation}
 \Pi=-\frac{\zeta}{\tau}\quad \rm{and} \quad\Phi=\frac{4\eta}{3\tau}.
\end{equation}
The Navier-Stokes hydrodynamics is known to have
 instabilities and acausal behaviours\cite{lindblom,BRW-06}; second order theories removes such 
unphysical artifacts.

We use causal dissipative second order hydrodynamics of Isreal-Strewart\cite{Israel:1979wp} to study the 
expanding plasma in the fireball. In this theory we have evolution equations for $\Pi$ and $\Phi$ governed by 
their relaxation times $\tau_{\Pi}$ and $\tau_{\pi}$. 
We refer \cite{AM-DR:04,ROM:09R} for more details on the recent developments 
in the theory and its application to relativistic heavy ion collisions.

Under these assumptions, the set of equations (i.e., equation of motion (\ref{edot}) and relaxation equations 
for viscous terms) dictating the longitudinal expansion of the medium are given by\cite{AM07,BRW-06,Heinz05} 
\begin{eqnarray}
   \frac{\partial\varepsilon}{\partial\tau} &=& - \frac{1}{\tau}(\varepsilon 
  + P +\Pi - \Phi) \, ,
  \label{evo1} \\
  \frac{\partial\Phi}{\partial\tau} &=& - \frac{\Phi}{\tau_{\pi}}+\frac{2}{3}\frac{1}{\beta_2\tau}
  -\left[ \frac{4\tau_{\pi}}{3\tau}\Phi +\frac{\lambda_1}{2\eta^2}\Phi^2\right] 
  \, , 
  \label{evo2} \\
   \frac{\partial\Pi}{\partial\tau} &=& - \frac{\Pi}{\tau_{\Pi}} - \frac{1}{\beta_0\tau} .
\label{evo3}
\end{eqnarray}
where $\Phi=\pi^{00}-\pi^{zz}$.
The terms in the square bracket in Equation(\ref{evo2}) 
are needed for the conformality of the 
theory\cite{Baier:2008:JHEP}. The coefficients $\beta_0$ and $\beta_2$ are related with the relaxation time by
\begin{equation}
 \tau_\Pi=\zeta\,\beta_0\,,\tau_\pi=2\eta\,\beta_2.
\end{equation}
We use the $\mathcal N\,=\,4$ supersymmetric Yang-Mills theory expressions for $\tau_\pi$ and $\lambda_1$
\cite{Baier:2008:JHEP,Okamura:N=4,shiraz:N=4}:
\begin{equation}\label{tau_pi}
  \tau_{\pi} = \frac{2-\ln 2}{2\pi T} 
\end{equation}
and 
\begin{equation}
 \lambda_1 = \frac{\eta}{2\pi T} \, .\label{lambda}
\end{equation}
We set $\tau_\pi(T)=\tau_\Pi(T)$ as we don't have any reliable prediction for $\tau_\Pi$\cite{fms08}.

In order to close the hydrodynamical evolution equations (\ref{evo1} - \ref{evo3}) we need to supply the equation of state.

\subsection{Equation of state, $\zeta/s$ and $\eta/s$}

We are interested in the effect of bulk viscosity on the hydrodynamical evolution of the plasma 
and recent studies show that near the critical temperature $T_c$ effect of bulk viscosity becomes 
important\cite{karsch07,Kharzeev:2007wb}. 
We use the recent lattice QCD result of A. Bazavov $\it{et ~al.}$\cite{Bazavov:2009} for equilibrium equation of state 
(EoS) (\textit{non-ideal}: $\varepsilon-3P\neq 0$). 
Parametrised form of their result for trace anomaly is given by
\begin{equation}
\frac{\varepsilon-3P}{T^4} = \left(1-\frac{1}{\left[1+\exp\left(\frac{T-c_1}{c_2}\right)\right]^2}\right)
\left(\frac{d_2}{T^2}+\frac{d_4}{T^4}\right)\ ,
\label{e3pt4}
\end{equation}
where values of the coefficients are $d_2= 0.24$~GeV$^2$, $d_4=0.0054$~GeV$^4$, $c_1=0.2073$~GeV, 
and $c_2=0.0172$~GeV\cite{kr:2010cv}. Their calculations predict a cross over from QGP to hadron gas around .200-.180 GeV. 
We take critical temperature as .190 GeV throughout the analysis. 
The functional form of the pressure is given by \cite{Bazavov:2009}
\begin{equation}
\frac{P(T)}{T^4} - \frac{P(T_0)}{T_0^4} = \int_{T_0}^T dT' \,\frac{\varepsilon-3P}{T'^5}\ ,
\label{pt4}
\end{equation}
with $T_0~$= 50 MeV and $P(T_0)$ = 0 \cite{kr:2010cv}.

From Equations (\ref{e3pt4}) and (\ref{pt4}) we get $\varepsilon$ and $P$ in terms of $T$.

\begin{figure}
\includegraphics[width=8.5cm,height=6cm]{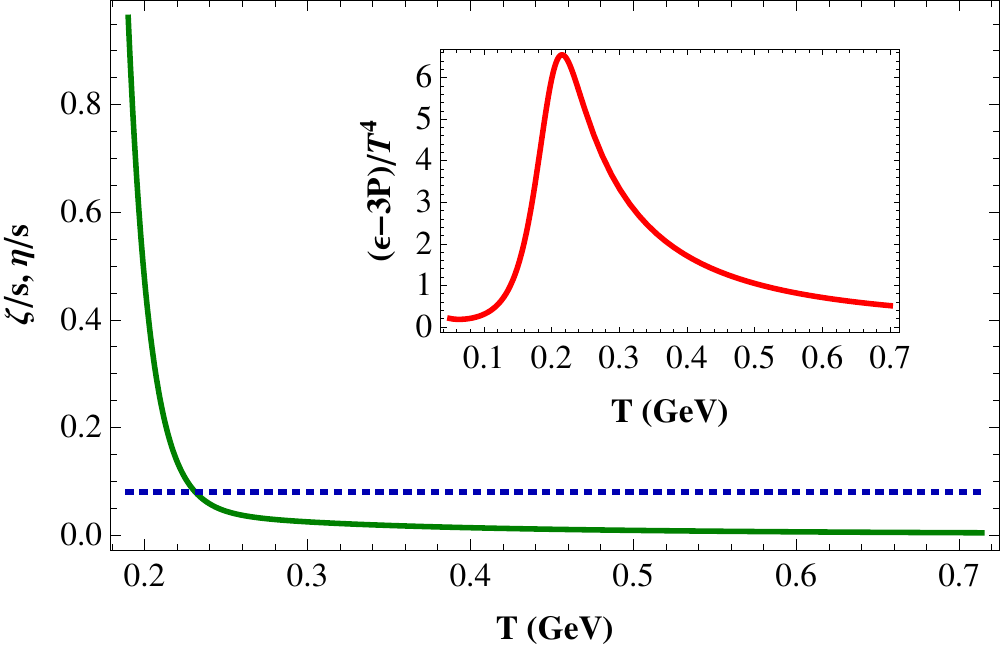}
\caption{$(\varepsilon-3P)/T^4\, ,\zeta/s$ (and $\eta/s=1/4\pi$) as functions of temperature T. 
One can see around critical temperature ($T_c=.190$ GeV)  $\zeta\gg\eta$ and departure of equation of state 
from ideal case is large.}
\label{fig1}
\end{figure}

\begin{figure}
\includegraphics[width=8.5cm,height=6cm]{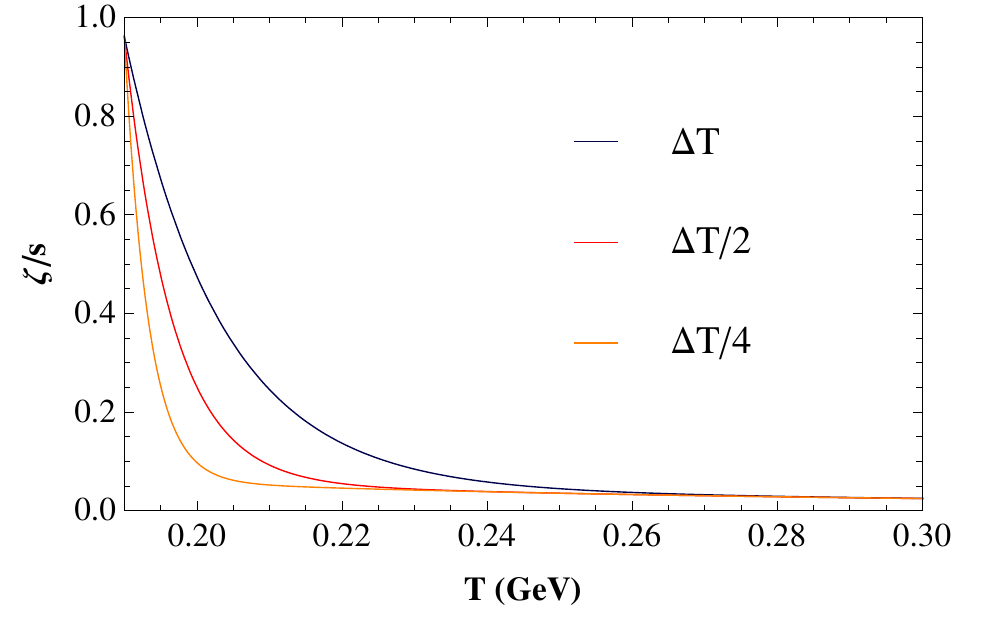}
\caption{Various bulk viscosity scenarios by changing the width of the curve through the parameter $\Delta T$.}
\label{fig1a}
\end{figure}

We rely upon the lattice QCD calculation results for determining $\zeta/s$. We use the result of 
Meyer\cite{Meyer:2008}, which indicate the existence a peak of $\zeta/s$ near $T_c$, however the height 
and width of this curve are not well understood. 
We follow parametrization of Meyer's 
result from Ref.\cite{kr:2010cv}, given by
\begin{equation}
\frac{\zeta}{s} = a \exp\left( \frac{T_c - T}{\Delta T} \right) + b \left(\frac{T_c}{T}\right)^2\quad{\rm for}\ T>T_c,
\label{zetabys}
\end{equation}
where $b$ = 0.061. The parameter $a$ controls the height and $\Delta T$ controls the width of the $\zeta/s$ curve 
and are given by
\begin{equation}
a=0.901,\, \Delta T=\frac{T_c}{14.5} .\label{zetacurve}
\end{equation}
We will change these values to explore the various cases of $\zeta/s$ to account for the uncertainty of the height 
and width of the curve. In FIG.\ref{fig1a} we show the change in bulk viscosity profile by varying the width of the 
$\zeta/s$ curve by keeping the height intact.

We use the lower bound of the shear viscosity to entropy density ratio known as KSS bound\cite{kss05} 
\begin{equation}
\eta/s=1/4\pi \label{KSS}
\end{equation}
in our calculations. We note that the entropy density is obtained from the relation
\begin{equation}
 s=\frac{\varepsilon+P}{T}\label{s}.
\end{equation}

In FIG.\ref{fig1} we plot the trace anomaly $(\varepsilon-3P)/T^4$ and $\zeta/s$ for desired temperature range. 
We also plot the constant value of $\eta/s=1/4\pi$ for a comparison. It is clear that the \textit{non-ideal} 
EoS deviates from the \textit{ideal} case ($\varepsilon=3P$) significantly around the critical temperature. 
Around same temperature $\zeta/s$ starts to dominate over $\eta/s$ significantly. We would like to note that these 
results are qualitatively in agreement with Ref.\cite{fms08}.

\subsection{Thermal photons}

During QGP phase thermal photons are 
originated from various sources, like \textit{Compton scattering} $q(\bar q)g\rightarrow q(\bar q)\gamma$ and annihilation 
processes $q\bar q\rightarrow g\gamma$. 
Recently Aurenche \textit{et al.} showed that two 
loop level \textit{bremsstrahlung} process contribution to photon production is as important as \textit{Compton} or 
\textit{annihilation} contributions evaluated up to one loop level\cite{aurenche:98}. They also discuss a new mechanism for hard 
photon production through the annihilation of an off-mass shell quark and an antiquark, 
with the off-mass shell quark coming from scattering 
with another quark or gluon. 
These processes in the context of hydrodynamics of heavy ion collisions were studied in 
Refs.\cite{dk99:EPJC,thoma07}. Until recently only the processes of 
\textit{Compton scattering} and \textit{$q\bar q$-annihilation} were considered in studying the photon production 
rates.

The production rate for hard ($E > T$) thermal photons from 
equilibrated QGP evaluated to the one loop order using perturbartive thermal QCD based on hard thermal loop (HTL) 
resummation to account medium effects. The \textit{Compton scattering} and \textit{$q\bar q$-annihilation} 
contribution is\cite{kapu91,baier92,traxler95}
 
\begin{equation}
 E\frac{dN}{d^4xd^3p}~=~\frac{1}{2\pi^2}\alpha\alpha_s \left(\sum_f e_f^2\right)~T^2~e^{-E/T}~\rm{ln}
\left(\frac{cE}{\alpha_s T}\right),\label{Compt+Ann}
\end{equation}
where the constant $c\approx$ 0.23 and $\alpha$ and $\alpha_s$ are the  electromagnetic and strong 
coupling constants respectively. 
In summation $f$ denotes the flavours of the quarks and $e_f$ is the electric charge of the 
quark in units of the charge of the electron.

The rate of photon production due to \textit{Bremsstrahlung} processes is given by\cite{aurenche:98}
\begin{equation}\label{Brems}
 E\frac{dN}{d^4xd^3p}~=~\frac{8}{\pi^5}\alpha\alpha_s \left(\sum_f e_f^2\right)~\frac{T^4}{E^2}~e^{-E/T}~(J_T-J_L)~I(E,T)
\end{equation}
where $J_T\approx1.11$ and $J_L\approx−1.06$ for two flavours and three colors of quarks\cite{thoma07}. 
The expression for $I(E,T)$ is given by
\begin{widetext}
\begin{equation}
 I(E,T)=\left[ 3\zeta(3) +\frac{\pi^2}{6}\frac{E}{T}+\left(\frac{E}{T}\right)^2 \rm{ln}(2) 
        + 4~\rm{Li}_3\left(-e^{-|E|/T}\right) + 2\left(\frac{E}{T}\right)\rm{Li}_2\left(-e^{-|E|/T}\right)
        - \left(\frac{E}{T}\right)^2~\rm{ln}\left(1+e^{-|E|/T}\right)\right]
\end{equation}
\end{widetext}
 
and $\rm{Li}$ are the polylogarithmic functions given by
\begin{equation*}
\rm{Li_a}(z)=\sum_{n=1}^{+\infty}\frac{z^n}{n^a}.
\end{equation*}
Now the rate due to $q\bar q$\textit{-annihilation with an additional scattering in the medium} is given by,
\begin{equation}\label{A+S}
 E\frac{dN}{d^4xd^3p}~=~\frac{8}{3\pi^5}\alpha\alpha_s \left(\sum_f e_f^2\right)~E~T~e^{-E/T}~(J_T-J_L).
\end{equation}

We use the parametrization of $\alpha_s(T)$ by Karsch\cite{karsch88}:
\begin{equation}
 \alpha_s(T)=\frac{6\pi}{(33-2N_f)~\rm{ln(8T/T_c)}}
\end{equation}
for our rate calculations. Here $N_f$ is the number of quark flavors in consideration.\\

\begin{figure}
\includegraphics[width=8.5cm,height=6cm]{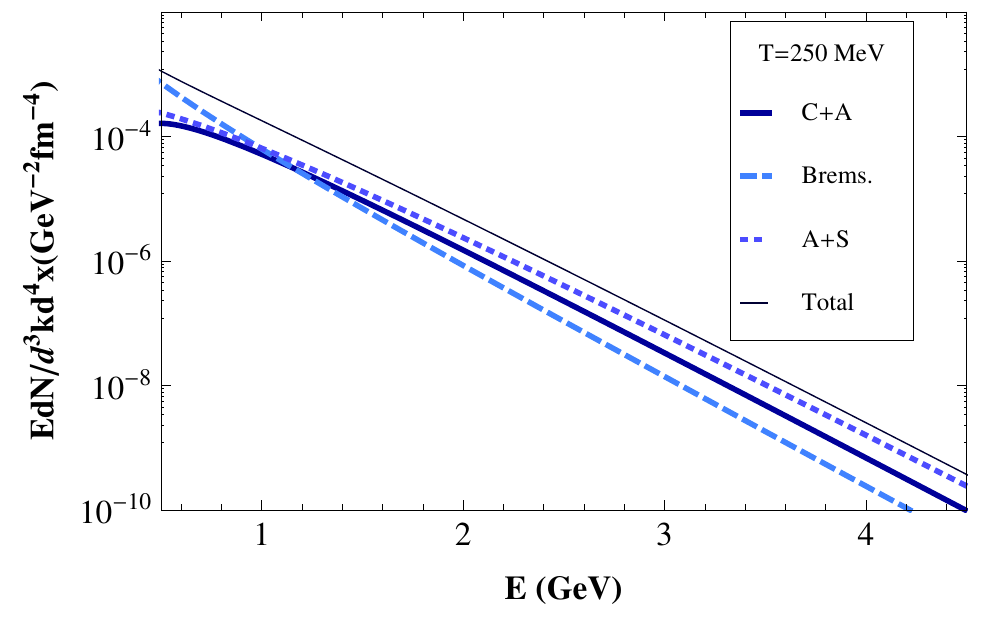}
\caption{Hard thermal photon rates in QGP as a function of energy for a fixed temperature T=250 MeV. Photon rates are 
plotted for different relevant processes.}
\label{fig2}
\end{figure}
In Fig.\ref{fig2}, we plot the different photon rates for a fixed temperature $T=250~MeV$. It shows the contributions from 
\textit{Bremsstrahlung} (Brems), \textit{annihilation with scattering} (A+S) and \textit{Compton scattering} together with 
\textit{$q\bar q$-annihilation} (C+A). \textit{Bremsstrahlung} contributes to the photon production rate 
upto $E\sim 1~GeV$ only, afterwards A+S and C+A processes become dominant. This observation is in complete agreement with 
with Ref.\cite{thoma07}.\\

The total photon rate is obtained by adding different temperature depended photon rate expressions.
Once the evolution of temperature is known from the hydrodynamical model, the \textit{total photon spectrum} 
is obtained by integrating the total rate over the space time history of the collision\cite{thoma},
\begin{eqnarray}\label{tot-yield}
\left(\frac{dN}{d^2p_\bot dy}\right)_{y,p_\bot}&=&\int d^4x\left(E\frac{dN}{d^3pd^4x}\right)\\
&=&Q\int_{\tau_0}^{\tau_1}
d\tau ~\tau \int_{-y_{nuc}}^{y_{nuc}}dy^{'}
\left(E\frac{dN}{d^3pd^4x}\right)\nonumber
\end{eqnarray}
where $\tau_0$ and $\tau_1$ are the initial and final values of time we are interested. 
$y_{nuc}$ is the rapidity of the nuclei whereas $Q$ is its transverse cross-section. 
For a $Au$ nucleus $Q \sim 180 fm^2$. $p_\bot$ is the photon momentum
in direction perpendicular to the collision axis.
The quantity $\left(E\frac{dN}{d^3pd^4x}\right)$ is  Lorentz invariant
and it is evaluated in the local rest frame in equation (\ref{tot-yield}). 
Now the photon energy in this frame, i.e., in the frame comoving with
the plasma, is given as $p_\bot cosh(y-y^\prime)$. So once the rapidity and $p_\bot$ are given we get the 
total photon spectrum.\\

\section{RESULTS AND DISCUSSION}

\begin{table}
\caption{Initial conditions for RHIC}
\vskip 0.1 in
\begin{center}
\begin{tabular}{ccc}
\hline

\multicolumn{1}{c}{$y_{nuc}$}&
\multicolumn{1}{c}{$\tau_0$} &
\multicolumn{1}{c}{$\rm{T_0}$} \\
\hline
\hline
$~$   &$(fm/c)$ &$(GeV)$\\
\hline
5.3   &0.5  &.310 \\
\hline
\end{tabular}
\end{center}
\end{table}

In order to understand the temporal evolution of temperature $T(\tau)$, pressure $P(\tau)$ and viscous stresses
- $\Phi(\tau)~\rm{and}~\Pi(\tau)$, we numerically solve the hydrodynamical equations describing the longitudinal expansion 
of the plasma: (\ref{evo1}-\ref{evo3}). We use the \textit{non-ideal} EoS obtained from equations 
(\ref{e3pt4}) and (\ref{pt4}). Information about viscosity coefficients $\zeta$ and $\eta$ are obtained from equations 
(\ref{zetabys}-\ref{KSS}) using equation (\ref{s}). 
We need to specify the initial conditions to solve the hydrodynamical equations, 
namely the initial time $\tau_0$ and $T_0$. 
We use the initial values relevant for RHIC experiment given in Table I, taken from Ref.\cite{dk99:EPJC}. 
We will take initial values of viscous contributions as $\Phi(\tau_0)=0$ and $\Pi(\tau_0)=0$. 
We would like to note that 
our hydrodynamical results are in complete agreement with that of Ref.\cite{kr:2010cv}.\\

Once we get the temperature profile we calculate the photon production rates. 
Total photon spectrum $E\frac{dN}{d^3pd^4x}$ (as a function of rapidity, $y$ and transverse momentum 
of photon, $p_\bot$) is obtained by adding different photon rates using 
equations (\ref{Compt+Ann}),(\ref{Brems}),(\ref{A+S}) and convoluting with the space time evolution of 
the heavy-ion collision with equation (\ref{tot-yield}). The final value of time $\tau_1$ is the time at which temperature 
evolves to critical value $\tau_f$, i.e.; $T(\tau_1)=T_c$. In all calculations we will consider the photon 
production in mid-rapidity region ($y=0$) only.\\

We will be exploring various values of viscosity and its effect on the system. 
Since there is an ambiguity regarding the height and width of $\zeta/s$ curve, we will vary the parameters 
$a~\rm{and}~\Delta T$ from its base value given in equation (\ref{zetacurve}). By this we will able to study the effect of 
variation of $\zeta$ on the system. The varied values of the parameters are represented by  $a'~\rm{and}~\Delta T'$. 
We note that unless specified we will be using the base values of bulk viscosity parameters (\ref{zetacurve}) in our 
calculations. 
Throughout the analysis we will keep the shear viscosity $\eta$ to its base value given by equation (\ref{KSS}). \\

In order to understand the effect of \textit{non-ideal} EoS in hydrodynamical evolution and subsequent photon spectra we 
compare these results with that of an \textit{ideal} EoS ($\varepsilon=3P$).
We consider the EoS of a relativistic gas of massless quarks and gluons. The pressure of such a system is given by
\begin{equation}
P=a~T^4\,;\,a=\left(16+\frac{21}{2}N_f\right)~\frac{\,\pi^2}{90}\label{idealP}
\end{equation}
where $N_f=2$ in our calculations. Hydrodynamical evolution equations of 
such an EoS within ideal (without viscous effects) Bjorken flow can be solved analytically and the 
temperature dependence is given by\cite{bjor}
\begin{equation}
T = T_0~\left(\frac{\tau_0}{\tau}\right)^{1/3} \label{idealT},
\end{equation}
where $\tau_0~\rm{and}~T_0$ are the initial time and temperature.
While considering the viscous effect of this \textit{ideal} EoS, we will solve the set of hydrodynamical equations 
(\ref{evo1} - \ref{evo2}), since effect of bulk viscosity can be neglected in the relativistic limit
when the equation of state $P=\varepsilon/3$ is obeyed \cite{wein}.\\

\subsection*{Hydrodynamics with \textit{non-ideal} and \textit{ideal} EoS }
\begin{figure}
\includegraphics[width=8.5cm,height=6cm]{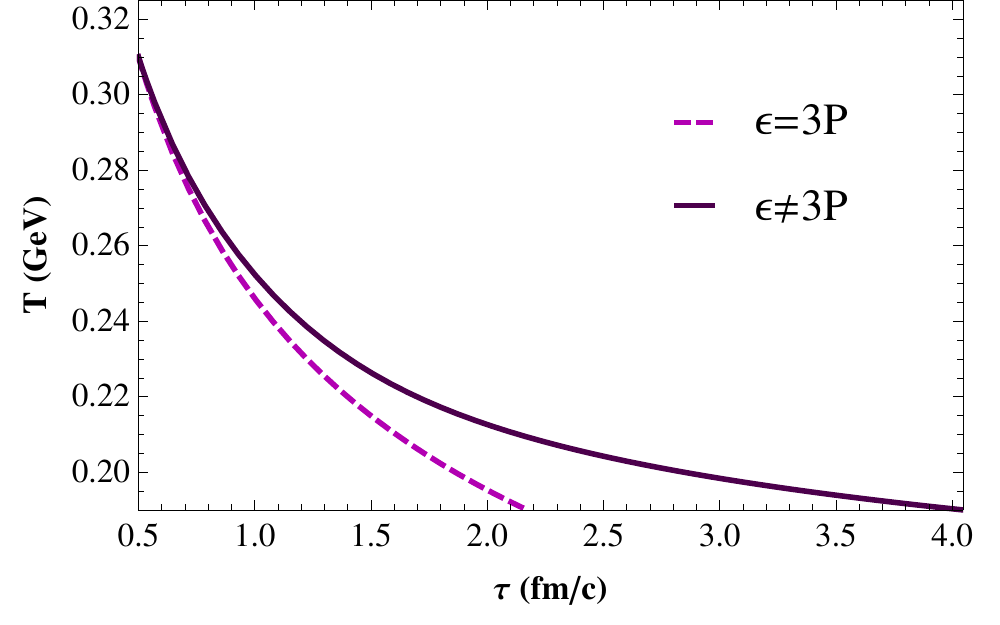}
\caption{Temperature profile using massless (\textit{ideal}) and \textit{non-ideal} EoS in RHIC scenario. 
Viscous effects are neglected in both cases. System evolving with \textit{non-ideal} EoS takes a significantly larger 
time to reach $T_c$ as compared to $ideal$ EoS scenario.}
\label{fig3}
\end{figure}
FIG. (\ref{fig3}) shows plots of temperature versus time for
the \textit{ideal} and \textit{non-ideal} equation of states. The temperature profiles are obtained
from the hydrodynamics without incorporating the effect of viscosity. The figure
shows system with \textit{non-ideal} EoS takes almost the double time than the system with
\textit{ideal} massless EoS to reach $T_c$. So even when the effect of viscosity is not considered, inclusion of 
the \textit{non-ideal} EoS makes significant change in temperature profile of the system. This can affect the corresponding 
photon production rates (below).\\

\begin{figure}
\includegraphics[width=8.5cm,height=6cm]{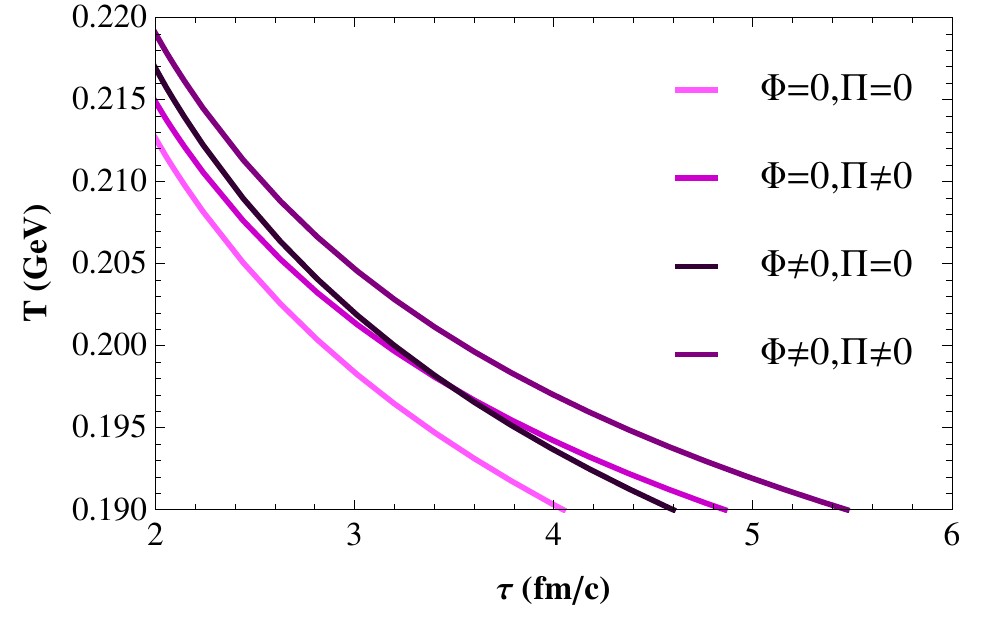}
\caption{Figure shows time evolution of temperature with \textit{non-ideal} EoS for different combinations of bulk 
($\Pi$) and shear ($\Phi$) viscosities. Non zero value of bulk viscosity refers to equations (\ref{zetabys}-\ref{zetacurve}) 
and non zero shear viscosity is calculated from equation (\ref{KSS}).}
\label{fig4}
\end{figure}
Now we analyse the viscous effects. Role of shear viscosity in the boost invariant hydrodynamics of heavy ion collisions, 
for a chemically nonequilibrated system, was already considered in Ref.\cite{skv-photon}. 

Next we consider possible 
combinations of $\Phi$ and $\Pi$ in \textit{non-ideal} EoS case and study the corresponding temperature 
profiles as shown in FIG. (\ref{fig4}). As expected viscous effects is slowing down temperature evolution. 
For the case of non zero bulk and shear viscosities ($\Pi\neq 0;~\Phi\neq 0$), temperature takes the longest 
time to reach $T_c$ as indicated by the top most curve. This is $\sim 1.5~ fm/c$ greater than the no viscosity case (the 
lowest curve). The remaining two curves show that the bulk viscosity  dominates over the shear
viscosity when the value of $T$  approaches $T_c$ and this makes the system to spend more time around $T_c$. 
However the intersection point of the two curves may vary 
with values of $a$ and $\Delta T$ as highlighted by FIG.\ref{fig1a}.

\subsection*{\textit{Non-ideal} EoS and Cavitation}

\begin{figure}
\includegraphics[width=8.5cm,height=6cm]{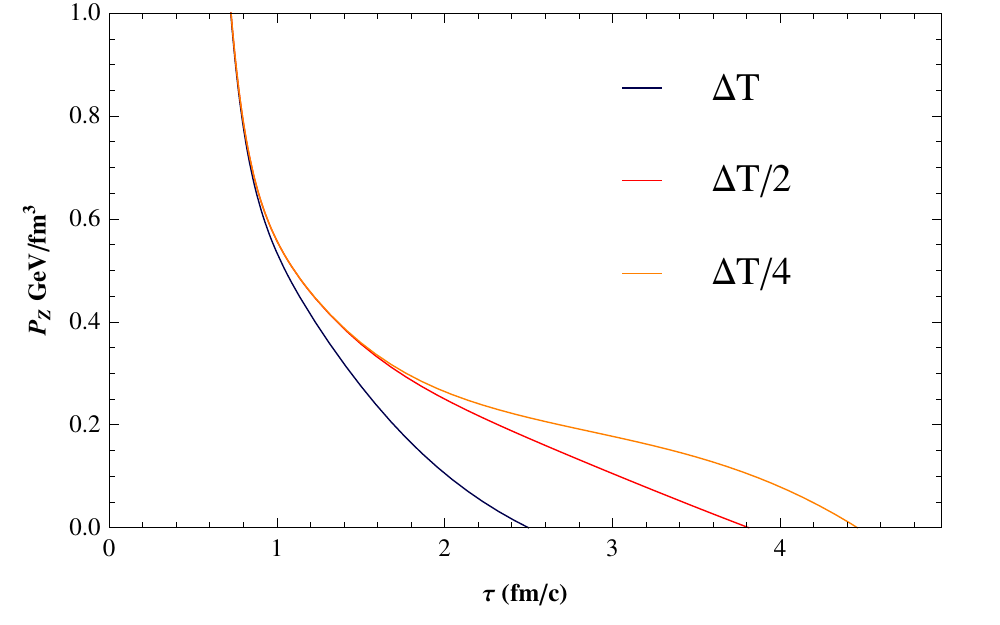}
\caption{Longitudinal pressure $P_z$ for various viscosity cases shown in FIG.\ref{fig1a}.}
\label{fig5}
\end{figure}

Let us note the fact that $\Pi<0$ \cite{kr:2010cv}. From the definition of longitudinal pressure $P_{z} = P + \Pi - \Phi$ 
it is clear that if either $\zeta$ ($\Pi$) or $\eta$ ($\Phi$) is large enough it can drive $P_z$ to negative values. 
$Pz=0$ defines the condition for the onset of \textit{cavitation}. At this instant when of $P_z$ becoming zero the 
expanding fluid will break apart in to fragments and hydrodynamic treatment looses its validity 
(see for e.g. Ref.\cite{kr:2010cv}). Recent experiments at RHIC suggest $\eta/s$ to its smallest value $\sim 1/4\pi$. 
And such a small value of $\eta/s$ alone is inadequate to induce cavitation. Therefor we vary the bulk viscosity values 
by changing $a~\rm{and}~\Delta T$ to study the cavitation. In the discussion that follows we will use $\tau_c$ to denote 
the time when cavitation occurs.\\
\begin{figure}
\includegraphics[width=8.5cm,height=6cm]{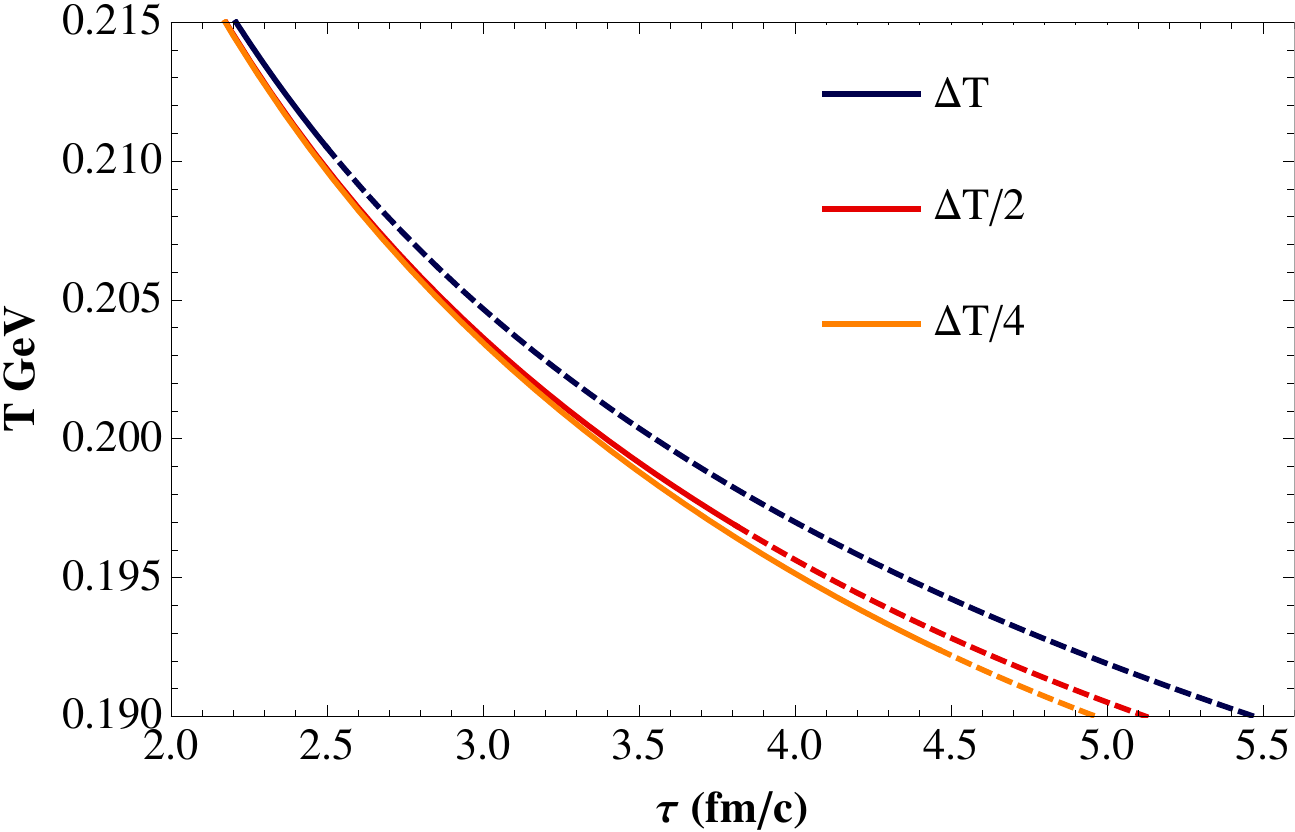}
\caption{Temperature is plotted as a function of time. With peak value ($a$) of $\zeta/s$ remains same while width 
($\Delta T$) 
varies. Solid line in the curve ends at the time of cavitation, while the dashed lines shows that how system would 
continue till $T_c$ if cavitation is ignored. Figure shows that larger the $\Delta T$ shorter the cavitation time.}
\label{fig7}
\end{figure}

In FIGs.\ref{fig5} and \ref{fig7} we plot $P_z$ and $T$ as functions of time for different values of $\Delta T$ while 
keeping $a$ (=0.901) fixed. It shows that higher value of $\Delta T$ 
is leading to a shorter cavitation time. For the values of $a~\rm{and}~\Delta T$ given by equation (\ref{zetacurve}) 
we find that around $\tau_c=2.5~fm/c$, $P_z$ becomes zero as shown by the curve at the bottom of the FIG.\ref{fig5}. 
In this case cavitation occurs when system temperature is larger than $T_c$. This can be seen from the top curve of the 
in FIG.\ref{fig7}. End point of the solid line in the top curve occurs at $T\sim 210~MeV$ and $\tau_c=2.5~fm/c$. 
Had we ignored the cavitation, system would have taken a time $\tau_f=5.5~fm/c$ to reach $T_c$ which is significantly 
larger than $\tau_c$ as seen from FIG.\ref{fig7}. This shows that cavitation occurs rather abruptly without giving any sign 
in the temperature profile of the system. The hydrodynamic evolution without calculating $P_z$ may end up in over 
estimating the evolution time and subsequent photon production.\\

\begin{figure}
\includegraphics[width=8.5cm,height=6cm]{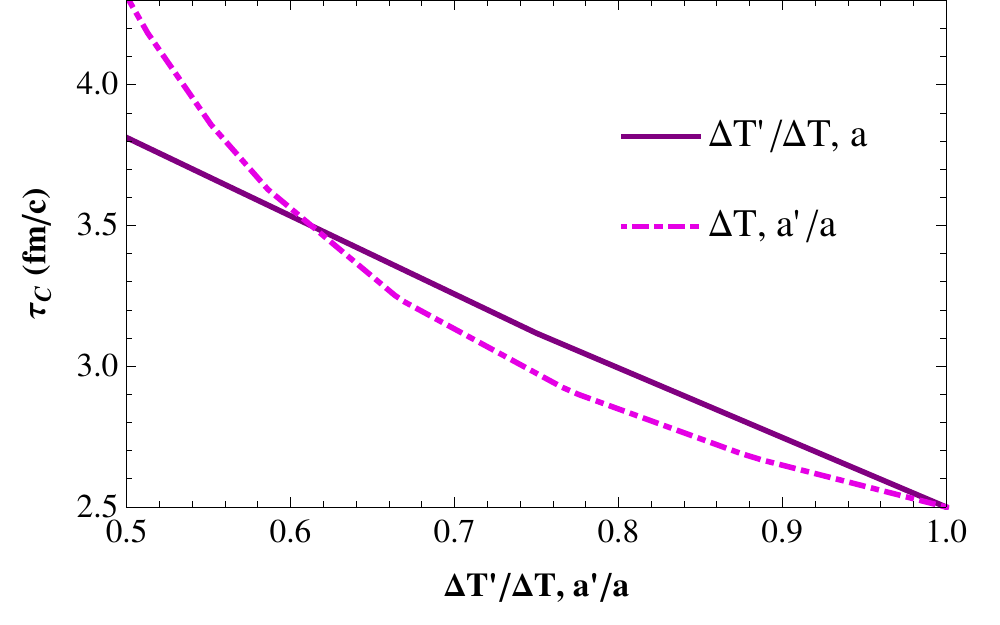}
\caption{Cavitation time $\tau_c$ as a function of different values of 
height ($a'$) and width ($\Delta T'$) of $\zeta/s$ curve.}
\label{fig8}
\end{figure}
A similar analysis can be carried out by keeping $\Delta T$ fixed ($=T_c/14.5$) and varying parameter $a$. 
We show the cavitation times corresponding to changes in $a$ and $\Delta T$ (denoted by $a'$ and $\Delta T'$) 
in FIG.\ref{fig8}. The dashed curve in FIG.\ref{fig8} shows $\tau_c$ as a function of $a$, while keeping $\Delta T$ fixed. 
The curve shows that $\tau_c$ decreases with with increasing $a$. Solid line shows how $\tau_c$ varies while keeping $a$ 
fixed and changing $\Delta T$.


\subsection*{Thermal Photon Production}
\begin{figure}
\includegraphics[width=8.5cm,height=6cm]{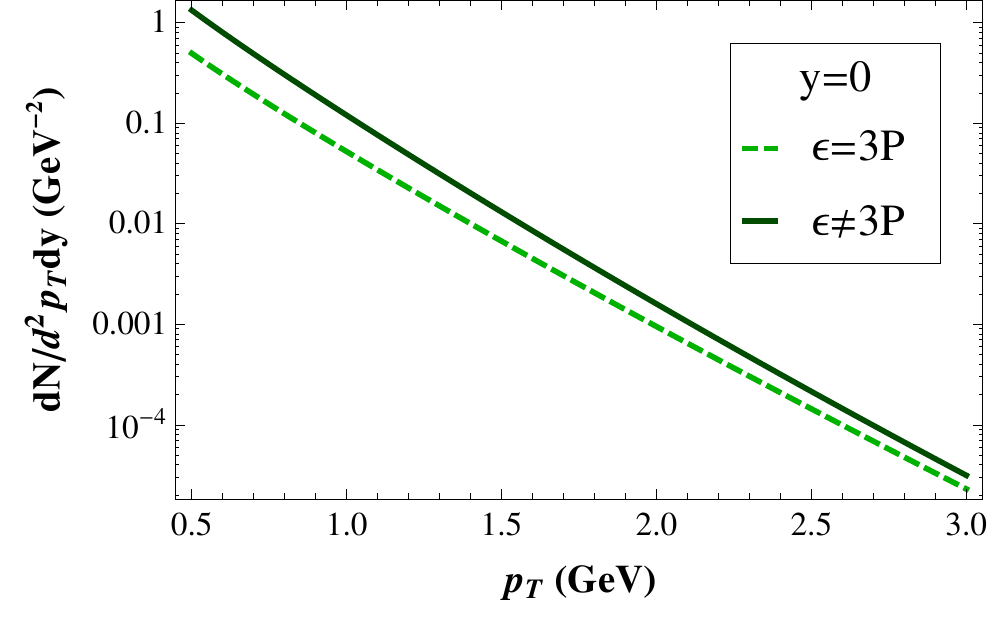}
\caption{Photon flux as function of transverse momentum for different equation of states.
No effect of viscosity included in the hydrodynamical equations.}
\label{fig9}
\end{figure}
We have already seen that the calculation of photon production rates require the initial time $\tau_0$, 
final time $\tau_1$ and $T(\tau)$. $\tau_1$ and $T(\tau)$ are determined from the hydrodynamics. Generally $\tau_1$ 
is taken as the time taken by the system to reach $T_c$, i.e.; $\tau_f$. 
But when there is cavitation, we must set $\tau_1=\tau_c$. Therefor 
photon productions will be influenced by cavitation, temperature profile and \textit{non-ideal} EoS near $T_c$.

\begin{figure}
\includegraphics[width=8.5cm,height=6cm]{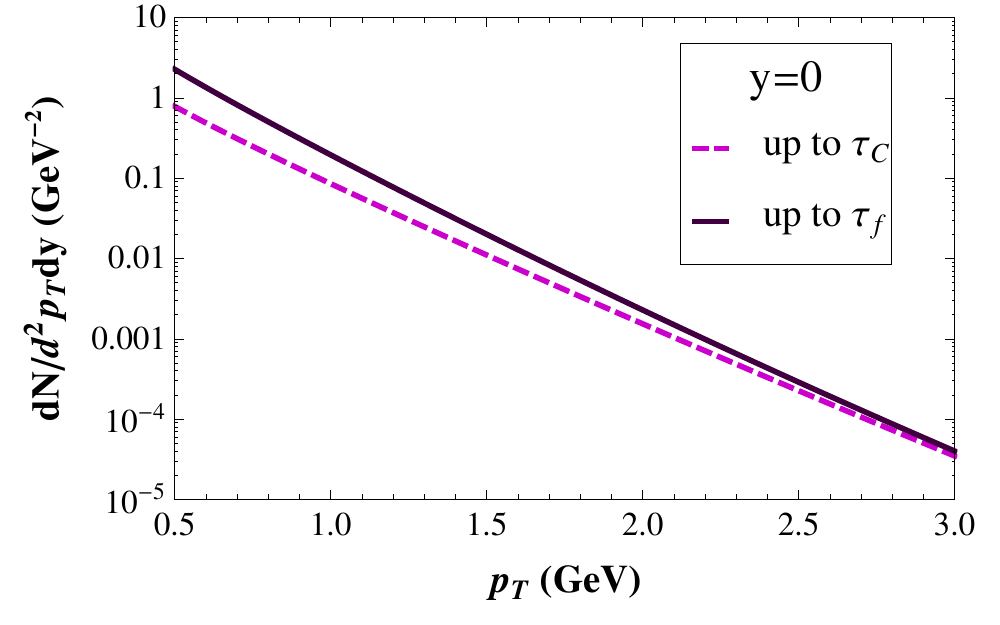}
\caption{Photon spectrum obtained by considering the effect of cavitation (dashed line). For a comparison we plot the 
spectrum without incorporating the effect of cavitation (solid line).}
\label{fig10}
\end{figure}
FIG. (\ref{fig9}) shows the photon production rate calculated using \textit{ideal} (massless) and \textit{non-ideal} EoS. 
The figure shows that \textit{non-ideal} EoS case can yield significantly larger photon flux
as compared to the \textit{ideal} EoS. At energy $E=1$ GeV, photon flux for the \textit{non-ideal} EoS is 60$\%$ larger 
than that of \textit{ideal} EoS case. This is because the calculation of the
photon flux is done by performing time integral over the interval between the initial time $\tau_0$ and
the final time $\tau_f$. $\tau_0$ is same for both the system while the
$\tau_f$ for the case with \textit{non-ideal} EoS is two times larger than the \textit{ideal} EoS.
Since the \textit{non-ideal} EoS allows the system to have consistently higher
temperature over a longer period as compared to the massless \textit{ideal-gas} EoS, more photons
are produced.

Next we try to observe the effect of cavitation in photon production. We emphasis that rates should only be integrated up 
to $\tau_c$. In FIG.\ref{fig10}, photon rates are calculated for the two cases. In the dashed curve the effect of 
cavitation is taken into account and $\tau_1=\tau_c=2.5~fm/c$. The solid line represent the same case but with the effect 
of the cavitation is ignored and $\tau_1=\tau_f=5.5~fm/c$. We see from the solid curve that we end up over estimating 
the photon rates at $p_\bot=0.5$ GeV by $\sim$ 200 $\%$ and $p_\bot=2$ GeV over estimation is about 50$\%$. 
So it is clear that information about cavitation time is crucial for correctly estimating thermal photon production rate.

\begin{figure}[h]
\includegraphics[width=8.5cm,height=6cm]{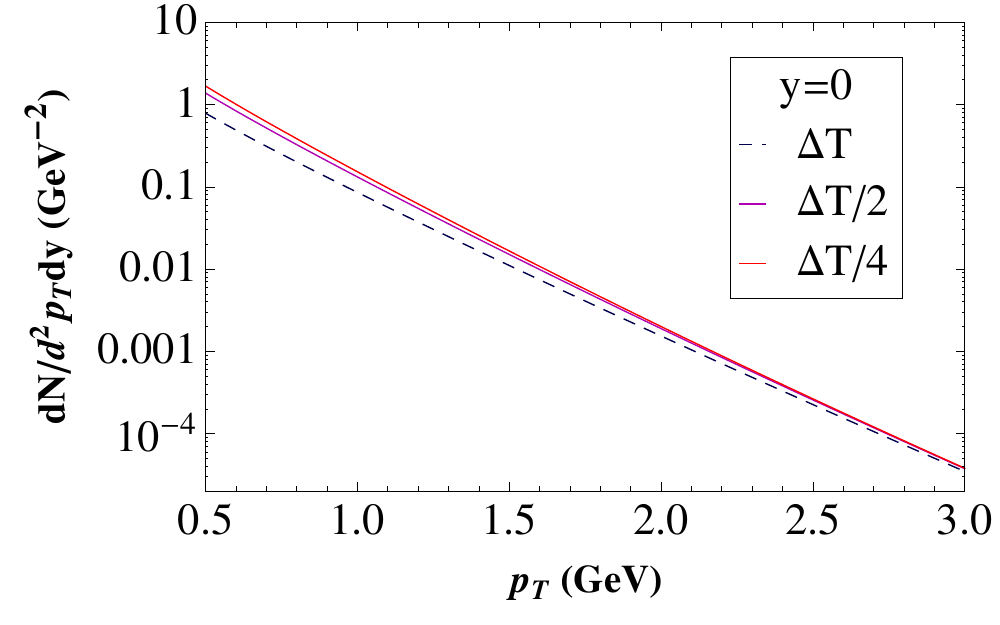}
\caption{Photon production rates showing the effect of different cavitation time.}
\label{fig11}
\end{figure}

In the FIG.\ref{fig11} we plot photon production rates for various cavitation times obtained by varying $\Delta T$ 
(with $a=0.901$ is fixed). Here the enhancement in the photon production when $\Delta T$ is reduced to half of its base 
value is $\sim75\%$ at $p_\bot=0.5$ GeV and $\sim 55\%$ at $p_\bot=1$ GeV. A further reduction of the parameter value 
to $\Delta T/4$ is enhancing the photon production by $\sim 110\%$ at $p_\bot=0.5$ GeV and $\sim 80\%$ at $p_\bot=1$ GeV. 
Reduction in $\Delta T$ amounts to increase in the cavitation time, which in turn would increase the time interval over 
which photon production is calculated. Therefor this increases the photon flux.

\section{SUMMARY AND CONCLUSIONS}
Thus using the second order relativistic hydrodynamics we have analyzed 
the role of non-ideal effects near $T_c$ arising due to the equation of
state, bulk-viscosity and cavitation on the thermal photon production.
Since the experiments at RHIC imply extremely small values for $\eta/s$,
the shear viscosity play a sub dominant role near $T_c$ in the photon production.

We have shown using non-ideal EoS using the recent lattice results that 
the hydrodynamical expansion gets significantly slow down as compared
to the case with the massless EoS. This in turn enhances  the flux of hard 
thermal photons. 

Bulk viscosity play a dual role in heavy-ion collisions: On one hand it enhances
the time by which the system attains the critical temperature, while on the
other hand it can make the hydrodynamical treatment invalid much before
it reaches $T_c$. We have shown that if the phenomenon of cavitation is ignored one
can have erroneous estimates of the photon production. Another result we would
like to emphasize is that reduction in cavitation time can lead
to significant reduction in the photon production. We hope that this
feature may be useful in investigating the signature of cavitation.


\end{document}